\documentclass{emulateapj}
\usepackage{graphicx}
\usepackage{natbib}
\shorttitle{The Clustering of Massive Halos}
\shortauthors{Wetzel et al.}

\begin{document}
\title{The Clustering of Massive Halos}
\author{Andrew R. Wetzel\altaffilmark{1,2}, J.D. Cohn\altaffilmark{1,3}, 
Martin White\altaffilmark{1,4}, Daniel E. Holz\altaffilmark{2,5},
and Michael S. Warren\altaffilmark{2}}
\altaffiltext{1}{Department of Astronomy, University of California, 
Berkeley, CA 94720}
\altaffiltext{2}{Theoretical Division, Los Alamos National Laboratory, 
Los Alamos, NM 87545}
\altaffiltext{3}{Space Sciences Laboratory, University of California, 
Berkeley, CA 94720}
\altaffiltext{4}{Department of Physics, University of California, 
Berkeley, CA 94720}
\altaffiltext{5}{Department of Astronomy \& Astrophysics, 
University of Chicago, Chicago, IL 60637}

\begin{abstract}
The clustering properties of dark matter halos are a firm prediction of
modern theories of structure formation.  We use two large volume,
high-resolution N-body simulations to study how the correlation function of
massive dark matter halos depends upon their mass, assembly history, and recent
merger activity.
We find that halos with the lowest concentrations are presently more clustered
than those of higher concentration, the size of the effect increasing with halo
mass; this agrees with trends found in studies of lower mass halos.
The clustering dependence on other characterizations of the full mass accretion
history appears weaker than the effect with concentration.
Using the integrated correlation function, marked correlation functions,
and a power-law fit to the correlation function, we find evidence that halos
which have recently undergone a major merger or a large mass gain have
slightly enhanced clustering relative to a randomly chosen population with
the same mass distribution.
\end{abstract}


\section{Introduction}

The observed Universe contains order on all scales we can probe.
It is generally believed that the largest structures arose via the
amplification of primordial (quantum) fluctuations during a period of
accelerated expansion, processed by the subsequent $13\,$Gyrs of
gravitational instability.
The pattern of clustering of objects on large scales is a calculable
prediction of cosmological models and thus comprises one of the
fundamental cosmological statistics.

Within modern theories of structure formation, the clustering of rare,
massive dark matter halos is enhanced relative to that of the general mass
distribution \citep{Kai84,Efs88,ColKai89,MoWhi96,SheTor99}, 
an effect known as bias.  The more massive the halo, the larger the bias.
As a result, the mass of halos hosting a given population of objects is
sometimes inferred by measuring their degree of clustering -- allowing a
statistical route to the notoriously difficult problem of measuring masses
of cosmological objects (e.g.~\citet{CooShe}).

Since halos of a given mass can differ in their formation history and
large-scale environment\footnote{
The large-scale environment of a halo refers to the density, smoothed
on some scale larger than the halos, e.g.~$5-10\,h^{-1}$Mpc.}, 
a natural question arises: do these details affect halo clustering?
In currently viable scenarios for structure formation, objects grow
either by accretion of smaller units or by major mergers with
comparable-sized objects.  
The formation history of a halo can thus be characterized by its mass 
accumulation over time, such as when it reached half of its mass,
had a mass jump in a short time, or last underwent a (major) merger.

Theoretically, the simplest descriptions of halo growth and clustering
\citep{bondetal,bower,lc93,lc94,kit-sut,kit-sut2}
do not give a dependence upon halo formation history
\citep{Whi93,SheTor04b,FurKam05,Har06}.  
To reprise these arguments: pick a random point in the universe and imagine
filtering the density field around it on a sequence of successively smaller
scales.  The enclosed density executes a random walk, which in the usual
prescription is taken to be uncorrelated from scale to scale.  The
formation of a halo of a given mass corresponds to the path passing a certain
critical value of the density, $\delta_c$, at a given scale.
The bias of the halo is the `past' of its random walk and its history the
`future' of the walk. 
All halos of the same mass at that time correspond to random walks crossing
the same point, and thus have the same bias. 
(Note that the derivation, using sharp $k$-space filtering, does not match
the way the prescription is usually applied, and this has been suggested by some
of the above authors as a way to obtain history dependence.  Introducing an
environmental dependence through e.g.~elliptical collapse will also give a
history dependence.)

The lack of dependence on halo history in the simplest descriptions does
not close the discussion theoretically or otherwise.
While these analytic methods work much better than might be expected given their
starting assumptions, the Press-Schechter based approaches still suffer many
known difficulties (e.g.~\citet{ShePit97}, \citet{BenKamHas05}).  
Other analytical ways of estimating the clustering of mergers have
been explored.  For example, \citet{FurKam05} defined a merger kernel 
(not calculable from first principles) and assumed that all peaks within
a certain volume eventually merged.  Such an ansatz implies that recently
merged halos are more clustered for $M>M_*$ and less clustered for $M<M_*$,
with some dependence upon predecessor mass ratios and redshifts. 
(Here $M_*(z)$ is the mass at which $\sigma(M)$, the variance of the linear
power spectrum smoothed on scale $M$, equals the threshold for linear density
collapse $\delta_c(z)$, see e.g.~\citet{Peacock}.)
Using close pairs as a proxy for recently merged halos, they found a similar
enhancement of clustering for $M>M_*$ and reduction for $M<M_*$ in several
(analytic) clustering models.  
To foreshadow our results: the signals we see are consistent  
with this trend.

Simple analytic models cannot be expected to capture all of the complexities
of halo formation in hierarchical models, and full numerical simulations are
required to validate and calibrate the fits.
Fortunately, numerical simulations are now able to produce samples with
sufficient statistics to test for the dependence of clustering on formation
history.  
Early work by \citet{LemKau99} showed that the properties of dark matter
halos, in particular formation times, are little affected by their
large-scale environment if the entire population of objects is averaged over.
They interpreted this as evidence against formation history and environment
affecting clustering.  As emphasized by \citet{SheTor04b}, however, this
finding --- plus the well known fact that the typical mass of halos depends
on local density --- implies that the clustering of halos of the same mass
must also depend on formation time.
Using a marked correlation function, \citet{SheTor04b} found that close pairs
tend to have earlier formation times than more distant pairs, work which was
extended and confirmed by \citet{Har06}.
\citet{GaoSprWhi05} found that later forming, low-mass halos are less 
clustered than typical halos of the same mass at the present; a possible
explanation of this result was given by \citet{WanMoJin06}.
\citet{Wec06} found a similar dependence upon halo formation time, showing
that the trend reversed for more massive halos and that the clustering
depended on halo concentration.  However, in order to probe to higher masses
these authors assumed that the mass dependence was purely a function of the
mass in units of the non-linear mass, then used earlier outputs to probe to
higher values of this ratio. 
It should be noted that scaling quantities by $M/M_*$ 
gives a direct equality only if clustering is self-similar.  Since $P(k)$ is not 
a power-law and $\Omega_{\rm mat}\ne 1$, a check of this approximation,
as is done here, is crucial.

These formation time dependencies are based on (usually smooth) fits to the
accretion history of the halo. 
However, halo assembly histories are often punctuated by large jumps from major
mergers that have dramatic effects on the halos.
Major mergers can be associated with a wide variety of phenomena, ranging from
quasar activity \citep{KauHae00} and starbursts in galaxies \citep{MihHer96} 
to radio halos and relics in galaxy clusters (see e.g.~\citet{Sar04} for
phenomena associated with galaxy cluster mergers).  Major mergers of galaxy
clusters are the most energetic events in the universe.
It follows that major merger phenomena can either provide signals of interest or
can cause noise in selection functions that depend upon a merger-affected
observable.  If recently merged halos cluster differently from the general
population (merger bias), and this is unaccounted for, conclusions drawn about
halos on the basis of their clustering would be suspect.  The question of
whether such merger bias exists remains unresolved, as previous work to
identify a merger bias through N-body simulations and analytic methods yields
mixed results \citep{Got02,Per03,ScaTha03,FurKam05}.

In this paper we consider the clustering of the most massive dark matter halos,
measured from two large volume ($1.1\,h^{-1}$Gpc)${}^3$ N-body simulations
described in \S\ref{sec:sims}.  We concentrate on massive halos, as most
previous simulations did not have the volume to effectively probe this end of
the mass function, and furthermore, for the largest mass halos the correspondence
between theory and observation is particularly clean.
We first examine the long-term growth history of halos, calculating the
``assembly bias'' as a function of growth history in \S\ref{sec:history},
extending previous results mentioned above to higher masses.
We then look to short-term history effects (i.e. events), measuring the 
``merger bias'' as a function of recent major merger activity or large mass
gain in \S\ref{sec:merger}, where we find a weak, but statistically significant,
signal for both cases. We conclude in \S\ref{sec:conclusions}.

\section{Simulations} \label{sec:sims}

To investigate the effects of formation history on clustering statistics
we use two high resolution N-body simulations performed with independent
codes: the HOT code \citep{HOT} and the TreePM code \citep{TreePM}. 
Both simulations evolved randomly generated, Gaussian initial conditions
for $1024^3$ particles of mass $10^{11}\,h^{-1}M_\odot$ from $z=34$ to
the present, using the same $\Lambda$CDM cosmology
($\Omega_M=0.3=1-\Omega_\Lambda$, $\Omega_B=0.046$, $h=0.7$, $n=1$ and
$\sigma_8=0.9$) in a periodic, cubical box of side $1.1\,h^{-1}$Gpc.
For the HOT simulation a Plummer law with softening
$35\,h^{-1}$kpc (comoving) was used.  The TreePM code used a spline
softened force with the same Plummer equivalent softening. 
The TreePM data were dumped in steps of light crossings of $136\,h^{-1}$Mpc
(comoving), producing 30 outputs from $z\approx 3$ to $z=0$.  
The HOT data were dumped from $z\approx 1$ (lookback time of 
$5.3\,h^{-1}$Gyr) to $z=0$ in intervals of $0.7\,h^{-1}$Gyr,
with the last interval at $z=0$ reduced to $0.4\,h^{-1}$Gyr. 
The outputs before $z\approx 1$ had so few high mass halos that the statistics
were not useful for the merger event calculations.
For comparisons of how using light crossings vs. fixed time steps in Gyrs
changes merger ratios, see \citet{CohWhi05}.
The TreePM simulations were used for the assembly histories and the HOT
simulations for the merger bias calculations -- though the results from the
two simulations were consistent so either could have been used in principle.

For each output we generate two catalogs of halos via the Friends-of-Friends
(FoF) algorithm \citep{DEFW}, using linking lengths $b=0.2$ and $0.15$ 
in units of the mean interparticle spacing.  These groups correspond roughly
to all particles above a density threshold $3/(2\pi b^3)$, thus both linking
lengths enclose primarily virialized material.
Henceforth halo masses are quoted as the sum of the particle masses within FoF
halos, thus a given halo's $b=0.15$ mass will be smaller than its $b=0.2$ mass
(see \citet{Whi01} for more discussion).
We consider halos with mass $M>5\times 10^{13}\,h^{-1}M_{\odot}$
(more than 500 particles);
at $z=0$ there are approximately $96,000$ such halos in each box for the
$b=0.15$ catalog and $120,000$ for the $b=0.2$ catalog.
The mass functions and merger statistics from the two 
simulations are consistent within Poisson scatter.

Given a child-parent relationship between halos at neighboring output times,
construction of the merger tree is straightforward since we are tracking
massive halos rather than e.g. subhalos.  
Progenitors are defined as those halos at an earlier time which
contributed at least half of their mass to a later (child) halo.
Of the approximately $10^5$ halos at $z=0$ we find only 14 for
which our simple method fails.  In these cases a ``fly-by'' collision of two
halos gives rise to a halo at $z=0$ with no apparent progenitors.
Excluding these halos does not change our results.
For the TreePM run, we use all 30 outputs to construct the merger tree,
which stored all of the halo information (mass, velocity dispersion, position,
etc.) for each halo at each output.  Each node of the tree pointed to a linked
list of its progenitors at the earlier time, enabling a traversal of the tree
to find mass accretion histories and mergers.  
The HOT run produced outputs for each time interval of child and
parent halos.

\section{Measuring clustering}

A basic measure of clustering is the two-point function, which in
configuration space is the correlation function, $\xi(r)$.
To compute $\xi(r)$ we use the method of \citet{LanSza92}:
\begin{equation}
\xi(r) = \frac{\langle DD\rangle-2\langle DR\rangle+\langle RR\rangle}
                {\langle RR\rangle} \; ,
\end{equation}
where $D$ and $R$ are data and random catalogs, respectively, and the angle
brackets refer to counts within a shell of small width having radius $r$.  
In computing $\langle DR\rangle$ and $\langle RR\rangle$ we use 
$10\times$ as many random as data points.  
To compute errors, we divide the simulation volume into 8 octants and compute
$\xi(r)$ within each octant. Since we probe scales much smaller than the
octants, we treat them as uncorrelated volumes, and we quote the mean $\xi(r)$
and error on the mean under this assumption.  
These errors tend to be $\sim1.4$--$2$ times larger than the more approximate
$\sqrt{N_{\rm pair}}$ error estimates used in some previous work. 
                                   
Our goal is to test the dependence of the clustering of objects
associated with some history dependent property.
A relevant quantity for comparison is the (mass dependent) bias of the halos
relative to the underlying dark matter, which we define as:
\begin{equation}
  \xi(r) = b^2 \xi_{\rm dm}(r).
\end{equation}

Analytically, the large-scale bias is related to a derivative of the halo mass
function \citep{Efs88,ColKai89,MoWhi96,SheTor99}.  
For the Sheth-Tormen form of the mass function one finds
\begin{equation}
  b_{ST}(M_{180 \rho_b}) = 1 + \frac{\nu'^2-1}{\delta_c}
       + \frac{0.6}{\delta_c\left(1+\nu'^{0.6}\right)},
\end{equation}
where $\nu' =0.841\delta_c/\sigma(M_{180 \rho_b})$ and $\delta_c = 1.686$.
This has been improved upon using the Hubble volume simulations
\citep{Col00,Ham01} --- see also \citet{SelWar04} 
for discussion of the bias defined through $P(k)$ on similar scales.
\citet{Ham01} used FoF halos with $b=0.164$ and found
\begin{eqnarray}
  b(M, R, z) &=& b_{ST}(M_{108},z) \nonumber \\
  &\times& \left[1.0 + b_{ST}(M_{108}, z)\sigma_R(R, z)\right]^{0.15} .
\end{eqnarray} 
The subscripts on the mass $M$ indicate which overdensity threshold is being
used to define the halo mass.
We took $M=0.93\,M_{108}$ and $M=1.07\,M_{180b}$, calculating the
conversion using the profile of \citet{NFW} assuming a concentration $c=5$.
The change in conversion factor was less than a percent for the range of
concentrations of interest.
See \citet{Whi01} for more details, discussion and definitions.

\begin{figure}
\begin{center}
\resizebox{3.1in}{!}{\includegraphics{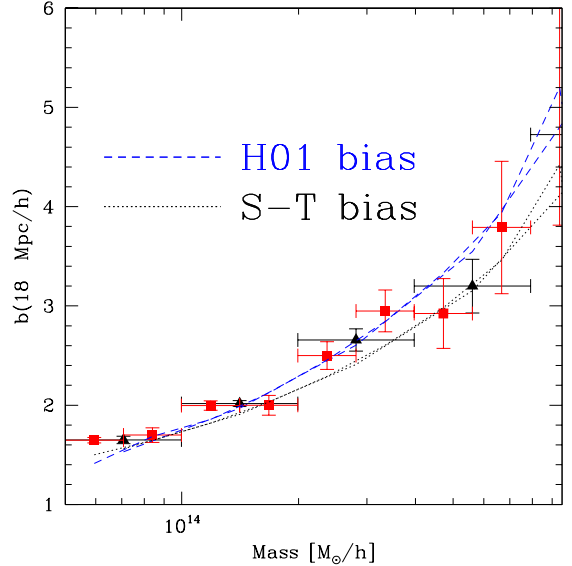}}
\end{center}
\caption{The bias $b(r)=\sqrt{\xi(r)/\xi_{\rm dm}(r)}$ at 
$r=18\,h^{-1}$Mpc for two different binnings in mass.  The horizontal
error bars on each point show the range of masses used.  The bias was
approximately scale-invariant in this mass regime from $15-30\,h^{-1}$Mpc.
We show 2 fits to $b(M)$ proposed in the literature: that of
\protect\citet{Ham01} (dashed) and
\protect\citet{SheTor99} (dotted),
each plotted for both mass binnings.}
\label{fig:hbias}
\end{figure}

We show the bias $b=\sqrt{\xi(r)/\xi_{\rm dm}(r)}$ at $r=18\,h^{-1}$Mpc 
as a function of mass in Fig.~\ref{fig:hbias}. 
The bias for halos with $M > 5\times 10^{13}\,h^{-1} M_\odot$ changed
less than 5\% on scales $r\ge 15\,h^{-1}$Mpc. 
We include the two bias fits given above for $r=18\,h^{-1}$Mpc at $z=0$.
The \citet{Ham01} fit was derived from a larger simulation volume; 
Fig.~\ref{fig:hbias} is included to illustrate the mass dependence of
the global bias, to provide a comparison context for the sizes of the
additional biases of concern in this paper.
We now turn to estimates of bias effects due to the history of the halos.

\begin{figure*}
\begin{center}
\resizebox{6.2in}{!}{\includegraphics{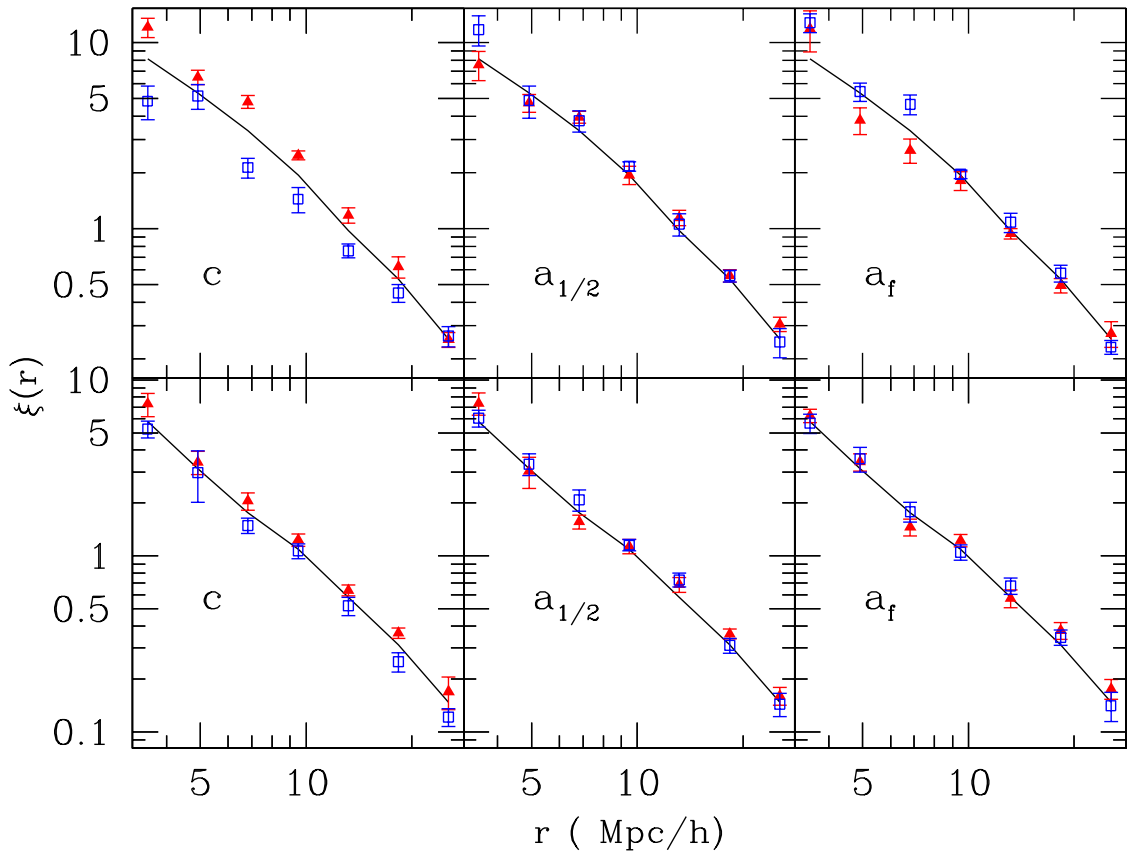}}
\end{center}
\caption{Correlation function of the lowest (filled triangles) and
highest (open squares) quartiles of (reduced) concentration,
$c$ (left),
half mass scale factor, $a_{1/2}$ (center) and formation scale factor,
$a_f$ (right).  The solid line is $\xi(r)$ for the full halo
sample.  Top panels: 
$10^{14}\,h^{-1}M_\odot\leq M\leq 3\times 10^{14}\,h^{-1}M_\odot$ 
(31551 halos). Bottom panels: $5\times 10^{13}\,h^{-1}M_\odot\leq 
M\leq 8\times10^{13}\,h^{-1}M_\odot$ (43638 halos).   
A clear signal is seen for concentration and formation scale factor
for the more massive halos.}
\label{fig:histcorr}
\end{figure*}

\section{Assembly bias} \label{sec:history}

We begin by considering parameterizations of the formation history of halos
which emphasize the global properties, i.e.~those related to the halo mass
growth over a long period of time.  
We consider three parameterizations of halo histories which have previously
been used with lower mass halos: $c$, $a_{1/2}$, and $a_{f}$ 
\citep{Wec06,GaoSprWhi05,SheTor04b}.  
Using these parameterizations
\citet{SheTor04b}, \citet{Har06}, \citet{GaoSprWhi05},
\citet{Wec06}, and \citet{CroGaoWhi06}
have shown that the clustering of halos of fixed mass is correlated with
``formation time'', a result which has come to be termed assembly bias.
The effect is strongest for smaller halos, and this has been the focus of
earlier work.
For the extremely massive halos that we consider halo identification
 is simpler, as none of our halos are subhalos.  
However, since massive halos are rarer, the statistics are poor even for a
simulation volume as large as ours. 

The concentration, $c$, is a parameter in an NFW fit to a halo density
profile \citep{NFW}\footnote{
We follow NFW and take $c=r_{200}/r_s$; note that \citet{Wec06} use 
$c_{\rm vir}=r_{\rm vir}/r_s$ where $r_{\rm vir}\simeq r_{100}$ for our
cosmology.  At $z=0$, $c_{\rm vir}\simeq 1.25\,c$.}.
We perform a least squares fit of the NFW functional form to the radial mass
distribution of all the particles in the FoF group, allowing $c$ and $M_{200}$
to vary simultaneously.
This is in order to be similar to the procedure of \citet{Bul01} to allow ready
comparison.
The concentration is expected to correlate with the time by which most of the
halo formed (earlier forming halos are more concentrated, see 
\citet{NFW96,Wec02,Gao04}).  
There is also a weak dependence of concentration on halo mass.
We have tried to minimize this effect by dividing out the average concentration
for each mass (calculated from the data) to get a ``reduced'' concentration,
which is essentially uncorrelated with mass (correlation is less than 0.2\%).

The second parameter encapsulating the formation history is $a_{1/2}$, 
the scale factor at which a halo accumulates half of its final mass.  
We find $a_{1/2}$ by linearly interpolating between the two bracketing times.
Analytic properties of this definition have been studied in \citet{SheTor04b},
and $a_{1/2}$ is often used as a proxy for formation epoch.
The third parameter, $a_f$, the formation scale factor, is also a formation 
time proxy.  It is defined through a fit to the halo mass accretion history
\citep{Wec02}\footnote{\citet{Mil06} present an analytic justification for
this form based on extended Press-Schechter theory.}:
\begin{equation}
  M(z)=M_0 \exp\left[ -2a_f z\right],
\end{equation}
where $M_0$ is the mass of the halo at $z=0$.
We calculate this from the history by doing a least squares fit
of $\ln(M_i/M_0)$ against $z_i$ for all the $z_i$ steps.
Although this form does not fit the mass accretion history of massive halos
particularly well due to their frequent mergers, the fit is well defined
and, as will be shown below, $a_f$ nonetheless appears to be correlated
with clustering. 

The correlations\footnote{Defined as
$(\langle a b \rangle -\langle a\rangle\langle b\rangle)/
\sqrt{\langle(a-\langle a\rangle)^2\rangle
      \langle(b-\langle b\rangle)^2\rangle}$,
see e.g.~\citet{Lup93}.} 
for many of the above parameters were presented in \citet{CohWhi05}.  
Some of these correlations have been compared in different combinations in 
\citet{Wec02}, \citet{Zha03a}, \citet{Zha03b},
\citet{Wec06}, and \citet{CroGaoWhi06}.
Except for \citet{Zha03a,Zha03b}, these were for galaxy scale halos rather
than galaxy cluster scale halos.  
The formation histories for low mass halos tend to be smoother and better
fit to the form of \citet{Wec02}, since they undergo fewer mergers
than high mass halos at late times.
Wechsler and Zhao give a formula for the concentration in terms of the 
formation time of \citet{Wec02}; our correlation coefficient is characterizing 
the scatter around any such correlation.  
For the current sample the strongest correlation ($0.69$) is between the 
formation redshift, $z_f = 1/a_f -1$, and the half-mass redshift,
$z_{1/2}=1/a_{1/2}-1$, consistent with the 0.70 found by \citet{CohWhi05}
with a sample about 1/7 the size.  
The formation redshift, $z_f$, and reduced concentration have a correlation 
of $0.53$.  
The full concentration and $z_{1/2}~(z_f)$ have a correlation of $0.56~(0.54)$. 
These correlations increase as the lower mass limit is decreased from
$10^{14}\,h^{-1}M_\odot$ to $5 \times 10^{13}\,h^{-1}M_\odot$.

To highlight any effects of assembly bias we take the highest and lowest
quartiles of the distribution of each of these three parameterization values and
compare the resulting $\xi(r)$ to that of the full sample 
(similar to \citet{Wec06}).  
We show examples for 
$10^{14}\,h^{-1}M_\odot<M<3\times 10^{14}\,h^{-1}M_\odot$ and 
$5\times 10^{13}\,h^{-1}M_\odot<M<8\times 10^{13}\,h^{-1}M_\odot$
in Fig.~\ref{fig:histcorr}.
For the higher mass halos we see a strong dependence of clustering on
concentration.  
We see a similar, but noticeably smaller, dependence on $a_f$, indicating that
more recently formed objects cluster more strongly.
As all of the objects we consider have $M>M_*$, our results are in line with the
expectation of \citet{Wec06} and the theoretical model of
\citet{FurKam05}.  
Specifically, this confirms the result found by \citet{Wec06} at $z=0$,
without needing to make the approximation that $b(c,M,z)=b(c,M/M_*)$.

The ratio of their correlation function at their top $c$ quartile to the total
sample for halos $\sim 10\,M_*$ was $\sim 1.25$.  
This is larger than our ratio, which doesn't reach 1.2 for any of the radii
considered in Fig.~\ref{fig:histcorr}, though it is well within our (and their) errors.
This is mirrored for the lowest $c$ quartile where our effect is similarly
reduced but within the errors.
We are using reduced concentration, while they divide each halo's concentration
by the average concentration in its mass bin, $\tilde{c}_{\rm vir}$.
For the lower mass sample a much weaker trend is seen (e.g.~the ratios for
the quartiles when selected on concentration barely reaches 10\%), agreeing
with the expectation that the signal decreases as $M\to M_*$.
At fixed mass, the trend of $b$ with $c$ is consistent with the fit of 
\citet{Wec06}, but the trend is so weak relative to the noise 
that the result is of marginal significance.  

\citet{GaoSprWhi05} and \citet{Har06} found bias for $M>M_*$ based 
on $z_{1/2}$, where both the lowest and highest quartiles of $z_{1/2}$ tended to
be more clustered than the full sample.  
We see a hint of this as well, but the fluctuations are large.
\citet{CroGaoWhi06} also found more dependence of clustering
on $z_{1/2}$ (their formation time) than on concentration, 
once luminosity dependent bias was taken out.  Note that their luminosity 
dependence might include some of the history measured by concentration or 
$z_{1/2}$ and their focus was on galaxies populating the halos rather than
the halos themselves.

Note also that even though $z_f$ and $z_{1/2}$ are correlated, the correlation
is not strong enough so that bias in one implies bias in the other. 
The overlap of the upper and lower quartiles for these quantities for $M>
10^{14} \, h^{-1}M_\odot$ is 62\% and 54\% respectively.  
As the rest of the clusters differ, the overall biases can be quite different,
as seen in Fig.~\ref{fig:histcorr}.

Another formation time related quantity, the redshift of last mass jump by
20\% or more in a time step corresponding to the light crossing
time of $136\,h^{-1}$Mpc comoving, had correlations with $z_{1/2}~(0.70)$,
$z_f~(0.61)$, and $c~(0.40)$.
We found a small sign of bias in the correlation functions of its highest and
lowest quartiles as well, leading us to expect a merger bias signal, as will be
examined in \S\ref{sec:merger}.

In summary, we confirm and extend previous results to lower redshift
and higher mass for concentration dependent bias.  
We see a smaller signal for formation time bias, and we see very little (if any)
signal for bias based on when halos reach half of their mass.
Bias in concentration and half-mass redshift have been seen in previous
work for smaller masses at higher redshift; our results show a smaller bias,
but well within errors, at least for the concentration dependent bias.

\section{Merger Bias} \label{sec:merger}

In the previous section we demonstrated the dependence of $\xi(r)$ upon
halo formation history, characterized by an average property such as the
``formation time''.  As halo assembly histories are punctuated
by large jumps from major mergers, we can also ask whether the clustering
of recently merged halos differs from that of the general population.

Although the concept of a major merger is intuitively easy to understand,
there is no standard definition in the literature of ``merger'' or
``major merger'' (these terms will be used interchangeably henceforth).
In simulations, where the progenitors can be tracked and masses measured,
major mergers can be defined in terms of masses of the progenitors and the
final halo.  We define progenitors as those halos at an earlier time which
contributed at least half of their mass to a later halo at the time of interest. 
The three most common ways to define a halo merger are: (1) the mass
ratio of the two largest progenitors, $M_2/M_1<1$ (2) the same ratio, but
using the contributing mass of the two most mass-contributing progenitors, and  
(3) $M_f/M_i$, the ratio of the current halo mass to the total mass of its 
largest progenitor at an earlier time.
We also consider (4) $M_{f}/M_1$, the ratio of the current halo
mass to the largest contributed mass.
In our simulations the merger fraction per $0.7\,h^{-1}$Gyr with
$M_2/M_1>0.3$ increases by more than a factor of 3 from $z=0$ to 1.

\begin{figure}
\begin{center}
\resizebox{3.1in}{!}{\includegraphics{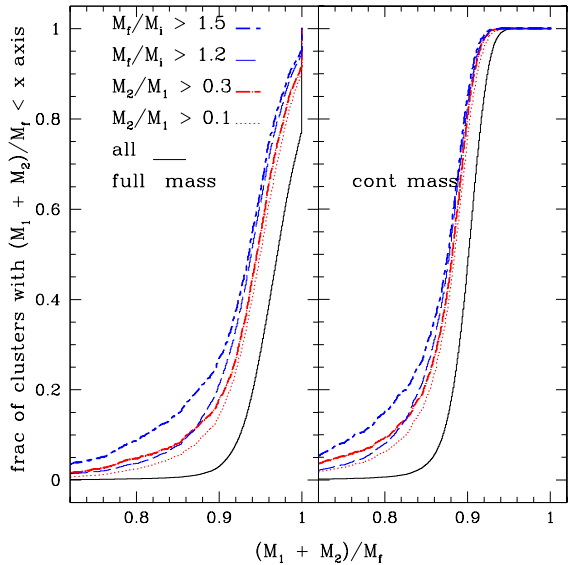}}
\end{center}
\caption{The cumulative distribution of $(M_1+M_2)/M_f$ for different
subsamples of our $b=0.15$ halos at $z=0$.  Looking back $0.4\,h^{-1}$Gyr
the subsamples are defined by $M_f/M_i>1.5$, 1.2 or $M_2/M_1>0.3$,0.1.
The lines are in the same order, top to bottom, with the lowest line being
the full sample.
At left $M_1$, $M_2$ are the full masses of the two largest progenitors,
at right $M_1$, $M_2$ refer to the contributing mass.}
\label{fig:mergedefs}
\end{figure}

One way to quantify how well the two body criteria ($M_2,M_1$ and $M_f,M_i$) 
describe the halo growth is to consider the ratio $(M_1+M_2)/M_f$.
This ratio is 1 for a halo formed only from its two largest
predecessors: a two body merger with no other accretion.  
It is lowered by accretion or multi-body mergers.
Fig.~\ref{fig:mergedefs} shows the cumulative distribution of
$(M_1+M_2)/M_f$ for halos with $M> 10^{14} h^{-1} M_\odot$ satisfying a
variety of merger criteria.
We considered both cases where $M_1$ and $M_2$ are the full and contributing
progenitor masses.  
As can be seen on the right, for all halos with $M>10^{14}\,h^{-1}M_\odot$ at
$z=0$, considering mass gains within the last $0.4\,h^{-1}$Gyr,
at least 5\% of the final halo mass is not from the two largest contributors.
As the merger criteria is hardened (i.e.~the merger is more ``major''), the
two largest progenitors contribute less and less of the final mass.
As can be seen on the left, the same amount of mass as found in the two largest
progenitors makes up the entire mass of the final halo in $\sim$25\% of the full
sample of halos. 
Lengthening the time step or looking to higher redshift also increases the
fraction of halos getting their mass from halos other than the two largest
progenitors.
For simplicity, our subsequent analysis uses only the two body criteria to
define mergers, so the accuracy of this assumption as examined above should be
kept in mind.

Previous work to identify a merger bias through N-body simulations and 
analytic methods gives a mixed picture.
\citet{Got02} found a clustering bias for recently ($\Delta t = 0.5\,$Gyr)
merged objects with $M_f/M_i>1.25$ and $M\leq M_*$ at $z=0$.
These authors, however, did not try to match the mass distribution of the
comparison sample to that of the merged halos --- a problem since mergers
occur more often for more massive halos, and the bias is known to increase
with halo mass.
To isolate the effects due to merging, the comparison sample needs to have
the same mass distribution as the merged sample, and most subsequent work
has ensured this.
\citet{Per03} found no bias between the correlation functions of recently
merged ($\Delta t=10^8\,$yr, $M_2/M_1>0.3$) and general samples at $z=2$
for halos with $M\sim M_*$, $25M_*$, and $150M_*$.
\citet{ScaTha03} confirmed Percival et al.'s results for major mergers in a
$z=3$ sample for a smaller range of masses, but surprisingly found an
enhancement of clustering for halos with recent 
($\Delta t=5\times 10^7$, $10^8\,$yr) large total mass gain, $M_f/M_i>1.20$. 
That is, they find a bias when selecting halos with a recent large mass gains,
but not when selecting on recently merged halos' parent masses. 
Their signal was weak due to limited statistics.  

That the previous literature is inconclusive is to be expected, given that the
effects of merger history upon clustering are small, and extremely difficult to
measure numerically.  We expect the largest signal when $M\gg M_*$, but this 
is where the number density of objects is smallest.  In addition, the most
extreme mergers are the rarest, increasing the shot-noise in the measurement of
$\xi(r)$.
If we include more common events, the ``merged'' and ``comparison'' samples
become more similar, washing out the signal of interest.
At higher redshift, the merger rate increases, thus the merged and comparison
samples have more overlap unless the merger ratio is increased, leading to
worse statistics.
To try to overcome these statistical effects, we use our very large samples of
simulated halos to search for a merger, or temporal, bias.

To define a ``recent major merger'' requires both a choice of threshold for
one of the merger ratios and a choice of time interval.
As we expect the halo crossing time to be $\sim0.7\,h^{-1}$Gyr,
(e.g.~\citet{TKGK}; \citet{GotKlyKra01};
\citet{RowThoKay04}), we expect that outputs at this separation or shorter are
small enough to catch recently merged halos while they are still ``unrelaxed''. 
That is, a ``recent merger'' might be expected to correspond to a dynamically
disturbed halo.

We consider the four merger criteria mentioned above, as well
as a wide range of samples and merger definitions.  
We used 9 different time intervals from $z\approx1$ to $z=0$ as given in
\S\ref{sec:sims}.  
We considered 4 different thresholds for both $M_2/M_1$ and $M_f/M_i$
using both total and contributing mass of the progenitors:
$M_2/M_1>0.1,0.2,0.3,0.5$ and $M_f/M_i>1.2,1.3,1.5,2.0$.  
Furthermore, we used two minimum masses, 
$5\times 10^{13}\,h^{-1}M_\odot$,
$10^{14}\,h^{-1}M_\odot$, and two FoF linking lengths, $b=0.15,0.2$.
Combinations of each of these criteria resulted in over 700 different pairs of
``merged'' and ``comparison'' samples.   
Although this data set is very rich, systematic trends are difficult to
identify.  This is in part because increasing the merger ``strength''
simultaneously increases the noise (due to lower numbers of events). 

Evidence of bias is very slight in the binned $\xi(r)$.
We used three methods to try to isolate the signal:
the marked correlation function, the integrated correlation function, and a
likelihood fit to a power law for the correlation function.
The clustering and merger criteria influence these three quantities in distinct
ways.
We now describe each method, and our corresponding results, in turn.

\begin{figure}
\begin{center}
\resizebox{3.1in}{!}{\includegraphics{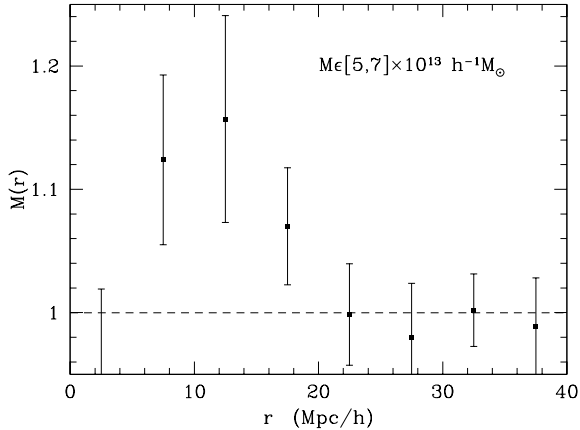}}
\end{center}
\caption{The marked correlation function for halos in the range
$5$--$7\times 10^{13}\,h^{-1}M_\odot$ at $z=0$.  The mark is the maximum
progenitor mass ratio, $M_2/M_1$, within the last $1~h^{-1}$Gyr.
The error bars come from dividing the sample into 8 octants.}
\label{fig:mark2}
\end{figure}

\subsection{Marked correlation function}

One problem with computing merger effects in terms of $\xi(r)$ is that, to
compute the difference in clustering of merged and random samples, one must
define a boolean merger criterion --- a halo is either in the merged sample or
not.  As halo histories are complex, a more nuanced measure of merger
clustering is useful, and this can be provided by using the marked correlation
function \citep{BeiKer00,BeiKerMec02,Got02,SheTor04b,Har06,SheConSki05}.
Each of $N$ objects gets assigned a mark, $m_i$, for $i=1,\dots,N$.
Denoting the separation of the pair $(i,j)$ by $r_{i,j}$, the marked
correlation function, $M(r)$, is defined by
\begin{equation}
  M(r) = \sum_{ij} \frac{m_i m_j }{ n(r)\bar{m}^2},  \;
\label{eqn:marked}
\end{equation}
where the sum is over all pairs of objects $(i,j)$ with separation $r_{ij}=r$,
$n(r)$ is the number of pairs, and the mean mark, $\bar{m}$, is calculated
over all objects in the sample.
The marked correlation function ``divides'' out the clustering of the average
sample, and thus a difference in clustering is detected for $M(r)\ne 1$.

\begin{figure*}
\begin{center}
\resizebox{6.2in}{!}{\includegraphics{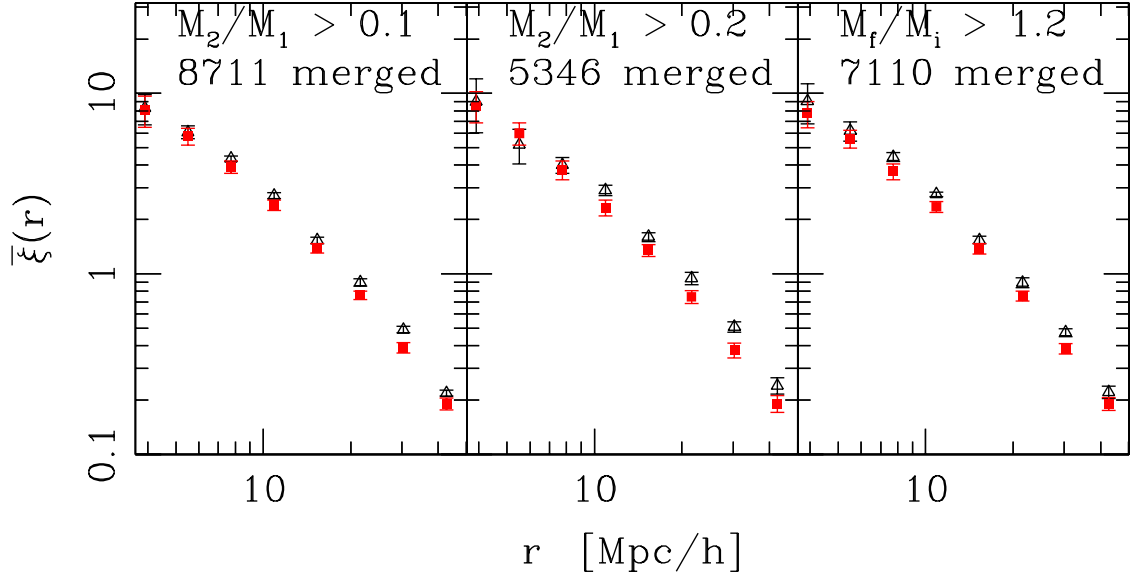}}
\end{center}
\caption{The integrated correlation function, $\bar{\xi}(r)$, of recently 
(within $0.4\,h^{-1} $Gyr) merged halos (triangles) and a comparison sample 
of the same mass (squares) for
$M_2/M_1>0.1$ (left), $M_2/M_1>0.2$ (middle), and $M_f/M_i>1.2$ (right),
where $M_1,M_2$ are the full masses of the progenitor halos, for 
halos in our $b=0.15$ catalog at $z=0$.  
The number of halos that merged out of the 96319 total halos with
$M> 5\times10^{13}\,h^{-1}M_\odot$ is shown at upper right in for each case.
For these three examples, the differences between the two samples are largest
at $30\,h^{-1}$Mpc, with significance $3.1\sigma$ (left), $2.7\sigma$ (middle),
and $2.5\sigma$ (right).}
\label{fig:xibars}
\end{figure*}

We consider five marks:
$M_2/M_1$ (for both total and contributed masses), $M_f/M_i$, $M_f/M_1$
(where $M_1$ is contributed mass) and $\frac{1}{2}(1 + M_2/M_1)$.
The last case had a smaller range of marks, and thus tests sensitivity to
extreme events.  The results for this mark were similar to the
others, suggesting that we are not dominated by outliers.
Halos are chosen with mass in a narrow range, 
$M_{\rm min}<M<\sqrt{2}M_{\rm min}$, to minimize the previously
mentioned bias due to merged halos being more massive.
The global bias changes less than a percent over the mass ranges we consider.

In our combined sample of several output times and mass ranges,
the largest signal comes from using as mark the maximum value
of $M_2/M_1$ within $\Delta t$ of the present, as shown in Fig.~\ref{fig:mark2}.
As $\Delta t$ was increased the signal went smoothly to zero.  
We find similar behavior for $M_2/(M_1+M_2)$, which suggests that any bias
is contributed by the systems where $M_2\ll M_1$. 
The signal is extremely weak for the other marks we considered.  
By stacking the signal across multiple output times (see
\S\ref{sec:lik} for details) we are able to find small, but
statistically significant detections of excess power for the
marks $M_2/M_1$, $M_2/(M_1+M_2)$, and $M_f/M_i$, for halos near 
$5\times10^{13}\,h^{-1}M_\odot$.
At higher masses there is weak evidence for an effect, but the large error bars
weaken the statistical significance.

As the marked correlation function approach finds only a weak signal,
typically an enhanced clustering of order 5--10\%,
we also explore two indicators which characterize the correlations by fewer
parameters: the integrated correlation function observed at a single scale,
and a likelihood fit to a power law correlation function.

\subsection{Integrated correlation function} \label{sec:xibar}

Given an object at some position, the integrated correlation function
\begin{equation}
  \bar{\xi}(r) \equiv \frac{3}{r^3}\int_0^r x^2\xi(x)\, dx  \; 
\end{equation}
is the probability, above random, that a second object will be within a sphere
of radius $r$.
This quantity enhances any increased clustering at short distances, but gives
error bars that are even more highly correlated than those of the correlation
function, $\xi(r)$, itself.  A typical result is shown in Fig. \ref{fig:xibars},
where a significant signal can be seen.
As in the previous section, we find a weak signal regardless of merger
definition in our 700 plus samples.  
Considering all the samples and all the separations $r$,
more than $2/3$ of the time the difference 
$\bar{\xi}_{\rm merge}(r)-\bar{\xi}_{\rm all}(r)$ was positive.

This method separates the data into radial bins, requiring us to
estimate the clustering at many locations.
Since the errors on the binned correlation points are highly correlated, we
reduced $\bar{\xi}(r)$ to a single measurement by fixing a preferred scale.
The signal tends to be largest near $r=20\,h^{-1}$Mpc (though the signal is
largest at $r=30\,h^{-1}$Mpc in the examples in Fig.~\ref{fig:xibars}), 
and so we compare $\bar{\xi}(r)$ of the merged and general samples at this 
radius.
On average, when a $2\sigma$ signal is seen (5--15\% of the time, depending
on mass ratio, etc.), $\bar{\xi}(r)$ for the mergers is $\sim$20\% higher
than for the general sample, although in extreme cases the difference can be
as large as a factor of 2 or 3.  
Due to the noisy statistics it was hard to identify any clear trends.

\subsection{Likelihood fit to $r_0$} \label{sec:lik}
\begin{figure*}
\begin{center}
\resizebox{6.2in}{!}{\includegraphics{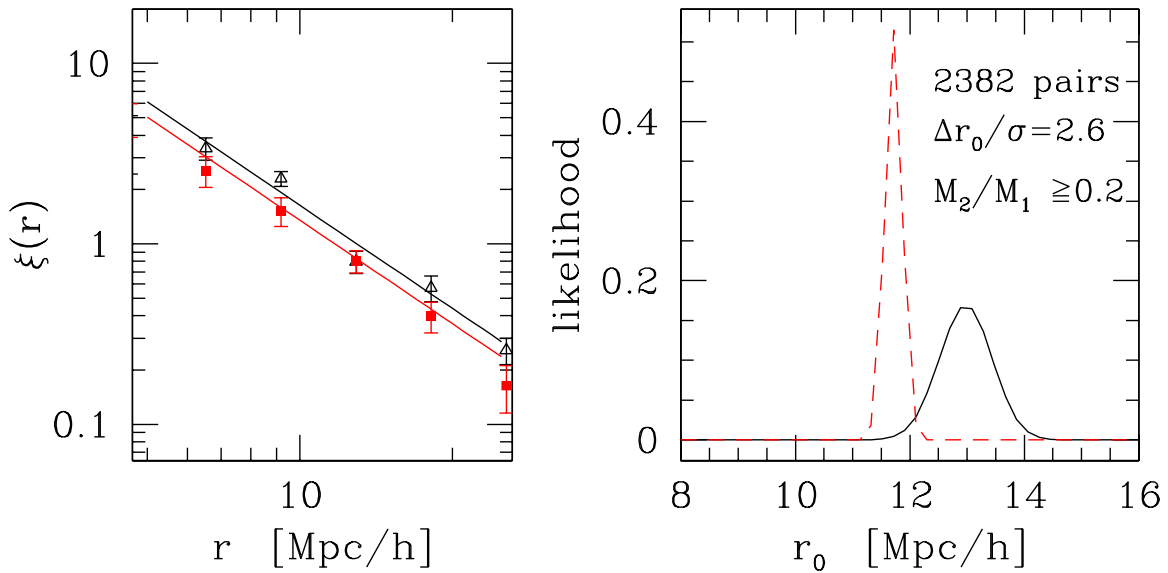}}
\end{center}
\caption{(Left) The correlation function for a recently merged sample
(triangles) and a comparison sample (squares) of the same mass.  The lines
indicate the best-fit $\gamma=1.9$ power law model, fit directly to the
cluster positions (not the binned $\xi(r)$).
(Right) The likelihood for the clustering amplitude, $r_0$, assuming a slope
$\gamma=1.9$ for the same samples at left. 
The sample is at $z=0$, with a minimum mass of 
$5\times 10^{13}\,h^{-1} M_\odot$ ($b=0.15$),
looking back $0.4\,h^{-1}$Gyr.  Mergers are tagged as having
$M_2/M_1 \geq 0.2$, $M_1,M_2$ full progenitor masses.}
\label{fig:maxl}
\end{figure*}

The integrated correlation function sums all pairs within a spherical region.
As an alternate approach, we approximate the correlation function as a power law
over some range of radii, and we perform a likelihood fit to this power law
correlation function:
\begin{equation}
  \xi(r) = \left(\frac{r}{r_0}\right)^{-\gamma}
\label{eqn:xipl}
\end{equation}
over the range of scales $(r_{\rm min},r_{\rm rmax})$.  
This method incorporates information from many scales, similar to the
integrated correlation function.  However, it is combined with the expectation
that the correlation function should be a power law and excises the center
region.  By using the positions of the halos directly in the fit
to the likelihood, the errors differ from those in the integrated
correlation function as well.

Assuming that the pair counts form a Poisson sample with mean proportional to
$1+\xi(r)$, the likelihood $L$ is \citep{Cro97,Ste97}
\begin{eqnarray}
  \ln L(r_0) &=&
  -2\pi\,\bar{n}^2 \int_{r_{\rm min}}^{r_{\rm max}} r^2
  \ \left[1 + \xi(r)\right]\,dr\nonumber \\
  &+& \sum_{i<j} \ln\left(\bar{n}^2r_{i,j}^2\left[1+\xi(r_{i,j})\right]\right)
  + {\rm const}\mbox{  ,}
\label{eqn:like}
\end{eqnarray}
where the sum is over measured pairs $i,j$ with separation $r_{i,j}$,
$\bar{n}$ is the measured average density\footnote{
We find that marginalizing or maximizing over $\bar{n}$ as a free
parameter results in biased fits for several samples.}, 
and $\xi(r)$ is given by Eq.~(\ref{eqn:xipl}).
We fit over the range $5$--$25\,h^{-1}$Mpc, where the correlation function
exhibits an approximately power law behavior.
For the comparison sample we multiply the likelihoods for several different
realizations, to reduce the noise, and then renormalize to unit area.
A typical result, where a significant signal can be seen, is shown in
Fig.~\ref{fig:maxl}, demonstrating both the power law fit and the maximum
likelihood distribution. For the fits, $r_0$ was usually $\sim 10\,h^{-1}$Mpc,
within the range where the power law fit was being applied.

\begin{figure}
\resizebox{3.1in}{!}{\includegraphics{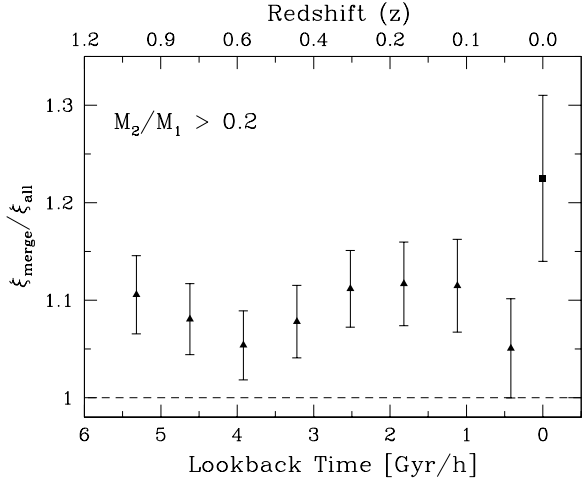}}
\caption{The (scale-independent) ratio of the power law fit correlation
 functions for the merged and comparison samples, as a function of lookback 
time/redshift; $\Delta t = 0.7\,h^{-1}$Gyr (triangles), $0.4\,h^{-1}$Gyr
(square).  The mergers satisfy the criterion $M_2/M_1>0.2$, with
$M_2,M_1$ total progenitor mass, for the 
$M>5\times 10^{13}\,h^{-1}M_\odot$ halos in our $b=0.15$ catalog. 
No evidence of systematic bias evolution with redshift is found. 
The enhanced clustering at $z=0$ arises presumably from the shorter
time interval used.}
\label{fig:biasevol}
\end{figure}

Across all of our samples, we find $\gamma\simeq 1.9\pm 0.1$.  
To allow us to compare different samples more easily, we reduce the number 
of free parameters to one by holding $\gamma\equiv 1.9$.
A typical example, demonstrating the ratio of the power law fit correlation
functions of the merged and general sample, is shown in 
Fig.~\ref{fig:biasevol} as a function of lookback time/redshift.
Since we fix $\gamma=1.9$ for both the merged and general sample,
the ratio $\xi_{\rm merge}/\xi_{\rm all}$ using Eq.~(\ref{eqn:xipl}) is
scale-invariant within our fit range. 
While the enhanced clustering of the recently merged sample is small, it
remains statistically significant.
Typically, the merged sample shows an enhanced clustering of 
$5-10\%$ in the correlation function for the $0.7\,h^{-1}$Gyr spacings, though
we find no strong evidence of systematic bias evolution with redshift.
Moreover, at $z=0$, where the spacing is smaller ($0.4\,h^{-1}$Gyr), we find a
significantly enhanced $\xi(r)$ for the mergers, often $10-20\%$.
Presumably, this increased clustering signal is caused by the smaller time interval.
Larger intervals encompass more mergers, leading to smaller errors, but also
leading to a smaller signal, since mergers now encompass a more
significant fraction of the comparison population.  As mentioned above,
looking at earlier times also makes the merged and comparison population overlap
increase dramatically.

\begin{figure}
\resizebox{3.1in}{!}{\includegraphics{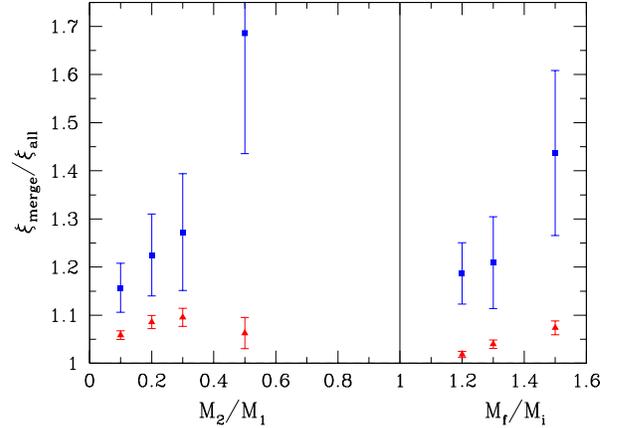}}
\caption{The (scale-independent) ratio of the power law fit correlation 
functions for the merged and comparison samples as a function of merger ratio, 
$M_2/M_1$ (left points; full progenitor mass) and $M_f/M_i$ (right points), 
for halos above $5\times 10^{13}\,h^{-1}M_\odot$ in our $b=0.15$ catalog.  
Mergers are counted within $0.4\,h^{-1}$Gyr of $z=0$ (squares), and an
average across all $0.7\,h^{-1}$Gyr spacings from $z\approx 1$ to
$z=0.04$ (triangles).
In both cases, clear trends can be seen.}
\label{fig:mratio}
\end{figure}

By averaging $\xi_{\rm merge}/\xi_{\rm all}$ across all of the $0.7\,h^{-1}$Gyr
spacings from $z\approx 1$ to $z=0.04$, we are able to study the size of the 
merger bias simply as a function of merger ratio.  
Figure \ref{fig:mratio} shows the increase of $\xi_{\rm merge}/\xi_{\rm all}$ 
with $M_2/M_1$ (full mass) and $M_f/M_i$ both for mergers within 
$0.4\,h^{-1}$Gyr of the present and for the redshift-averaged 
$0.7\,h^{-1}$Gyr spacings.
The merger bias clearly increases with increasing merger ratio, with the
smaller time step yielding stronger clustering as described above.

In summary, we find a weak bias in many cases (but not all---the signals are
very noisy) for recent major mergers and recent large mass gains.
While \citet{Per03} found no such merger bias, our signal is consistent
with their upper limit of $20\%$ on the bias effects of recent mergers.
The work of \citet{ScaTha03} saw a small bias for large mass gains but noted
that their statistics limited their ability to determine the significance.
Our larger box allowed us to incorporate the effects of cosmic variance, which
had been neglected in previous work.  
Cosmic variance increased the errors by 40\% or more, which limited the
significance of the signal.  
Nonetheless, we still found a small bias for \textit{both} mergers and large
mass gains.

\section{Conclusions} \label{sec:conclusions}

The large-scale structure of the Universe is built upon a skeleton of clustered
dark matter halos.  For the past two decades we have known that rarer, more
massive dark matter halos cluster more strongly than their lower mass
counterparts.
Halos of a fixed mass, however, can differ in their formation
history and large-scale environment, and recent work on halos
smaller than galaxy clusters has shown that this can lead to
further changes in their clustering.

In this paper we have used two large-volume, high resolution N-body simulations
to study the clustering of massive halos as a function of formation history.  
We confirmed earlier results that the lower concentration massive halos are
more clustered than the population as a whole;
extending these results to higher masses (and thus lower redshifts) 
than had been probed previously \citep{Wec06}.
(Previous work had looked at similar regimes of $M/M_*$ but for smaller $M$
and thus higher redshift; note again that exact scaling with $M/M_*$ is
not expected for non power law $P(k)$ and $\Omega_m \ne 1$.)  
Similarly, we confirmed the enhanced clustering of halos with later formation
times, though the signal was not as strong as for concentration.
The signal for bias based on a halo reaching half of its mass is weaker than
that seen in \citet{GaoSprWhi05} (again for higher $z$), and not statistically
significant in our case.  

We also investigated whether recent merger activity affected the clustering
of massive halos --- a topic with a muddied history in the literature.
While we found statistically significant ($>2\sigma$) merger effects on
clustering in many cases we considered, both for recent major mergers and
large mass gain, in most cases this signal was weak: a 5--10\% increase in bias.
Our strongest signal came from using a likelihood fit of the correlation 
function to a power law, particularly for major mergers within $0.4\,h^{-1}$Gyr
of the present, where we saw a typical merger bias of up to $20\%$.  
This bias signal is not necessarily at odds with the lack of signal in previous
work, which looked for larger bias than that seen on average here.  

Even with a $(1.1\,h^{-1}{\rm Gpc})^3$ volume, massive halos remain very
rare objects and small changes in their correlations are difficult to detect.
We were plagued by the competing effects that increasing the severity of the
merger (and hence underlying signal) decreases the number of pairs, worsening
the statistics.  
General trends remain elusive, since changing various criteria (e.g.~merger
definition, minimum mass, time step) generally changed the number of halos
involved, thus changing the errors.  
However, we did find that the strength of the merger bias typically increased
with increasing merger ratio, i.e.~more major mergers are more strongly biased.
Finally, we note that the correlations found between the last large (20\%)
mass gain and the different definitions of formation redshifts provide a
connection between the assembly bias studied in \S\ref{sec:history} and
the merger bias in \S\ref{sec:merger}.  
This bias is not expected from direct application of extended Press-Schechter
theory, and it provides a phenomenon that a more precise analytic model of
mergers should reproduce.

We thank J. Bullock, R. Croft, G. Jungman, P. Norberg, G. Rockefeller,
R. Sheth, E. Scannapieco, R. Wechsler and A. Zabludoff for enlightening
conversations and especially R. Sheth and R. Wechsler, who also provided
useful comments on the draft.
J.D.C., D.E.H. and M.W. thank the staff of the Aspen Center for Physics for
their hospitality while this work was being completed.
The simulations and analysis in this paper were carried out on supercomputers
at Los Alamos National Laboratory and NERSC.
J.D.C. was suported in part by the NSF.
M.W. was supported in part by NASA. D.E.H. gratefully acknowledges a
Feynman Fellowship from LANL.

\bibliographystyle{apj}
\end{document}